\documentclass[aps, prl, amsmath,amssymb, reprint]{revtex4-1}

\usepackage{graphicx}
\usepackage{dcolumn}
\usepackage{bm}
\usepackage[usenames, dvipsnames]{color}

\begin{document}

\preprint{AIP/123-QED}

\title[Magnetothermodynamics]{Magnetothermodynamics: Measuring equations of state in a relaxed MHD plasma}

\author{M. Kaur}
\email{mkaur2@swarthmore.edu.}

\author{L. J. Barbano}
\author{E. M. Suen-Lewis}
\author{J. E. Shrock}
\author {A. D. Light}
\affiliation{ 
Swarthmore College, Swarthmore, PA 19081 USA
}
\author{D. A. Schaffner}
\affiliation{%
Bryn Mawr College, Bryn Mawr, PA 19010 USA
}%
\author{M. R. Brown}
\affiliation{ 
Swarthmore College, Swarthmore, PA 19081 USA
}

\date{\today}

\begin{abstract}
We report the first measurements of equations of state of a fully relaxed magnetohydrodynamic (MHD) laboratory plasma. Parcels of magnetized plasma, called Taylor states, are formed in a coaxial magnetized plasma gun, and are allowed to relax and drift into a closed flux conserving volume. Density, ion temperature, and magnetic field are measured as a function of time as the Taylor states compress and heat. The theoretically predicted MHD and double adiabatic equations of state are compared to experimental measurements. We find that the MHD equation of state is inconsistent with our data.

\end{abstract}

\maketitle

Measuring the equations of state (EOS) of a compressed plasma is important both for advancing fusion experiments and understanding natural systems such as stellar winds. The true equation of state in the solar wind, for example, is still an open question. In the supersonic expansion of the magnetized solar wind, an anisotropy between perpendicular temperature ($T_\bot$) and parallel temperature ($T_\parallel$) with respect to the direction of the magnetic field is observed \cite{Kasper2003}. If expansion of the solar wind plasma were regulated by an adiabatic equation of state, one would expect the proton temperature (particularly $T_\bot$) to drop much faster with radius from the Sun than is measured \cite{Brown15}. The effects of collisional age \cite{Bale09} and kinetic instabilities \cite{Kasper02} appear to regulate the temperature anisotropy but do not yet account for the thermodynamics of the expansion. 

Over the past decades, numerous experiments \cite{park14,Hurricane14,Hammer88, Hammer91, Molvik91} motivated by fusion applications have been performed to achieve a highly compressed plasma but very little has been done to understand the thermodynamics of such systems. Some progress has been made in measuring the EOS of unmagnetized plasma in the context of inertial confinement fusion experiments \cite{park14,Hurricane14}. At the National Ignition Facility (NIF), a radial compression of unmagnetized DT pellets resulted in a density increase by a factor of 1000. Similar studies in magnetized plasma have yet to be performed.

Early compression experiments on magnetized plasma were performed at the RACE facility in the 1990s \cite{Hammer88, Hammer91, Molvik91}. In these experiments, spheromak-type plasmas were accelerated in coaxial electrodes to 2500 km/s and stagnated in a conical taper. In compression experiments on the IPA device, two field reversed configurations (FRCs) were merged and magnetically compressed to kilovolt ion temperature \cite{slough2011, Votroubek08}. In liner implosion experiments at the Shiva star facility, dense hydrogen plasma was compressed to megabar pressures \cite{Degnan99}. However, an equation of state (EOS) was not reported in any of these experiments. 

For the first time in laboratory experiments, we have explored directly the thermodynamics of compressed magnetized plasmas (``magnetothermodynamics"). In this Letter, we describe experiments in which we generate parcels of magnetized, relaxed plasma \cite{Taylor74, Woltjer58, cothran_prl, gray_prl} and observe their compression against a conducting cylinder closed at one end. The plasma parameters are measured during compression and a PV diagram is constructed to identify instances of associated ion heating. The local magnetohydrodynamic (MHD) and the double adiabatic (CGL) equations of state are tested during compression events, under several experimental conditions, showing that MHD equation of state is inconsistent with our observations.

In the collisional regime, MHD plasma with an isotropic velocity distribution can be treated as an ideal gas and the adiabatic equation of state for such a plasma can be written as
\begin{equation}
\frac{\partial}{\partial t}\left(\frac{P}{n^{\gamma}}\right) = 0 \label{mhd_eos},
\end{equation}
where $P$ and $n$ are the plasma pressure ($P = nk_B T$) and density, respectively; $T$ is the plasma temperature. 
$\gamma$ is the ratio of specific heats and is given by 
$ \gamma = (f + 2)/f $ where $f$ is the number of microscopic degrees of freedom.  

If the collision rate is low (i.e., $ \omega_{ci}\tau > 1 $, where $ \omega_{ci} $ and $\tau$ are the ion cyclotron frequency and Coulomb collision time, respectively), the perpendicular ($P_\bot = nk_B T_\bot$) and parallel pressure ($P_\parallel= n k_B T_\parallel$) with respect to the direction of the magnetic field are not the same and the MHD equation of state no longer remains valid. To account for such a situation, Chew, Goldberger and Low proposed a modified set of adiabatic equations of state, known as double adiabatic or CGL equations of state \cite{CGL}:
\begin{align}
	&\frac{\partial}{\partial t}\left(\frac{P_\bot}{nB}\right) = 0 \label{<cgl_perp>}, \\
	&\frac{\partial}{\partial t}\left(\frac{P_\parallel B^2}{n^3}\right) = 0 \label{<cgl_par>}.
\end{align}
Eq. (\ref{<cgl_perp>}) is related to the constancy of the first adiabatic invariant $ \mu = W_\bot/B$ and Eq. (\ref{<cgl_par>}) is related to the constancy of the second adiabatic invariant, $ \mathcal{J} = v_\parallel L $; where  $W_\bot$, $v_\parallel$ and $L$ are the perpendicular kinetic energy with respect to magnetic field, parallel velocity with respect to magnetic field and the length of the field line, respectively.

\begin{figure*}[t]
\includegraphics{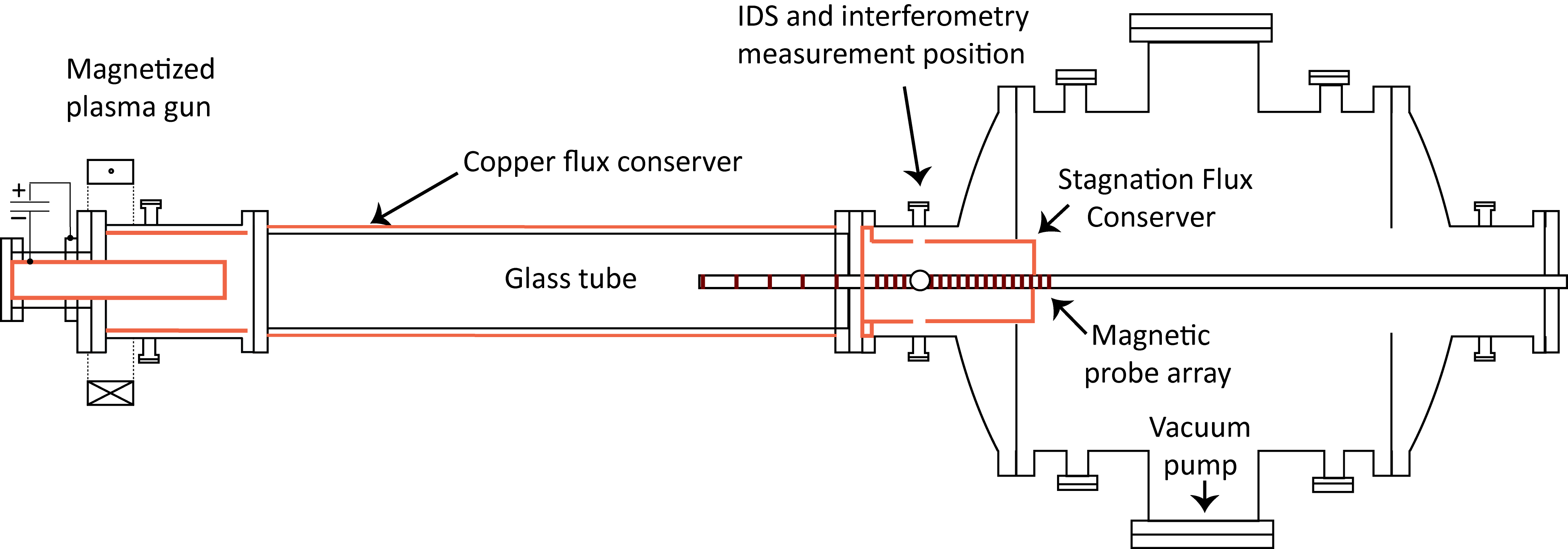}
\caption{\label{fig:schematic} A schematic of the experimental set-up. A glass tube has been added in between the gun and the stagnation flux conserver (SFC) and is covered with a copper flux conserving shell. All the three principal plasma diagnostics are located in the SFC at the location of the circle. $T_i$ is measured using ion Doppler spectroscopy along the vertical chord and $n$ is measured using HeNe laser interferometry along a horizontal chord. The long $\dot{B}$ probe array is aligned with the axis of the SFC.}
\end{figure*}

In these experiments, we produce parcels of magnetized plasma using a coaxial magnetized plasma gun located at one end of the linear Swarthmore Spheromak Experiment (SSX) device, as shown in Fig.~\ref{fig:schematic}.
The diameter of the inner electrode of the gun is $6.2~cm$ and the outer electrode diameter is $15~cm$. More details about the plasma gun can be found in earlier publications \cite{Geddes1998,Brown15,Brown_psst}. At the other end of the linear device, a closed, tungsten-lined copper can, referred to here as a stagnation flux conserver (SFC), is installed. The SFC is $30~cm$ long and has the same inner diameter as that of of the outer electrode. A $1~m$ long glass tube (diameter = $15~cm$) is installed in between the gun and the SFC, and is covered with a copper flux conserving shell having long magnetic soak time ($> 200~\mu s$). We maintain a good vacuum and regularly clean the gun and the device using He glow discharge to maintain a plasma with low impurity level.

The gun is initially prepared with a strong magnetic field ($\backsim 1~T$) in the inner electrode. Hydrogen gas is puffed into the annular region between the two electrodes. A voltage pulse ($\approx 4~kV$) is applied between the two electrodes which ionizes the gas and causes a high current ($\backsim 100~kA$) to flow through the plasma. $\mathbf{J} \times \mathbf{B}$ forces accelerate the plasma out of the gun and a toroidal self-consistent magnetic object, called a spheromak \cite{Geddes1998}, is formed. The toroidal structure continues to move away from the gun and simultaneously tilts and relaxes to a twisted minimum energy state, called a Taylor state; which is a force-free MHD equilibrium that has been well-characterized in previous studies \cite{Taylor74, Woltjer58, cothran_prl, gray_prl, Schaffner2014}. The inertia of the Taylor state carries it to the SFC, where it stagnates and compresses. 

For these experiments, we rely on three principal diagnostics. 
For ion temperature measurements, we use ion Doppler spectroscopy (IDS). For IDS studies, we make use of the nascent $C_{III}$ impurity ions present in our plasma \cite{Chaplin2009}. The $C_{III}$ line ($229.687~nm$) is collected along a vertical chord using a telescope, dispersed on an echelle grating (observed at $25^{th}$ order), and is recorded using a photomultiplier tube (PMT) array at $1~ MHz$ cadence \cite{cothran_rsi}. The line-averaged density of the Taylor state is measured along a horizontal chord using a $632.8~ nm$ HeNe laser interferometer. The interferometry chord lies in the same plane as the IDS chord.

Along with these two diagnostics, we use a long, axial $\dot{B}$ probe array to measure the Taylor state velocity using a time of flight (ToF) technique. The magnetics data is also used to determine the structure of the Taylor states and the local vector magnetic field co-located with the intersection of the IDS and interferometry chords. While selecting these diagnostics, our goal was to perform minimally invasive measurements. For assessing the equations of state, it is important that the measurements perturb the plasma as little as possible. Because of this, we have no way of measuring different profiles during compression.

\begin{figure}
	\includegraphics[width=0.48\textwidth]{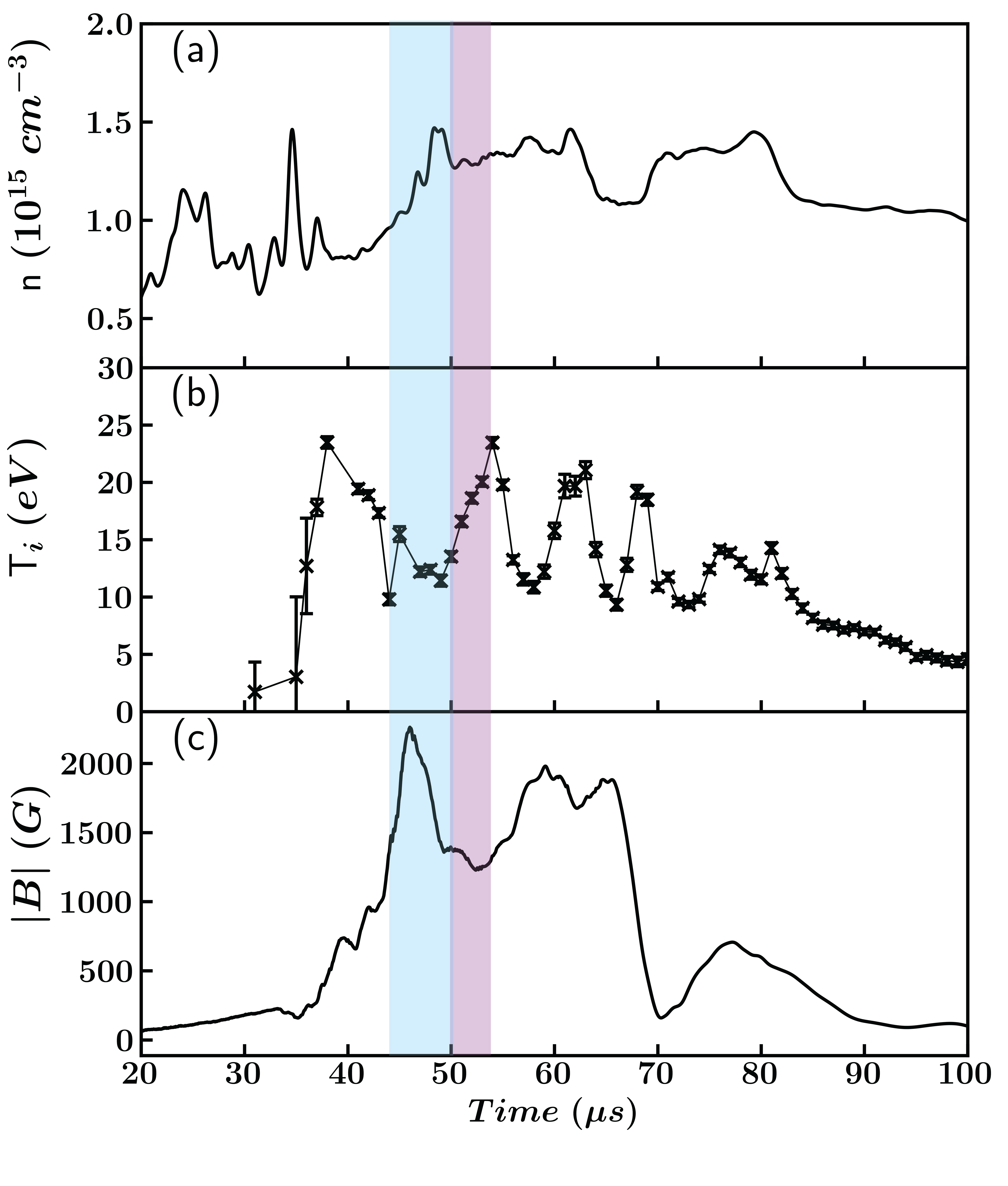}
	\caption{\label{fig:plasma_param} A typical time trace of a) plasma density, b) ion temperature and c) magnetic field measured inside the stagnation flux conserver. The error bars for $T_i$ are indicated by the vertical bars at each time value whereas the uncertainty in $n$ and $B$ is $\sim 10\%$.
	The Taylor state enters the SFC at $ 44~\mu s$ (indicated by the blue bar) and then compresses from $50  - 54~ \mu s$ against the closed end of the SFC accompanied by a rise in $T_i$ (indicated by the pink bar).}
\end{figure}

A typical set of time traces of plasma density, ion temperature and magnetic field measured inside the compression volume (SFC) is shown in Fig.~\ref{fig:plasma_param}%
. The Taylor state moves at $35 - 40 ~km/s$ and reaches the SFC at $50~\mu s$, indicated by the blue color in Fig.~\ref{fig:plasma_param}. 
We observe a rise in plasma density and ion heating for a time window ranging from $50 - 54~\mu s$, indicated by the pink color in Fig.~\ref{fig:plasma_param}, corresponding to compression against the back of the SFC. 
The Taylor state flow speed is consistent with free expansion and is not fast enough for the occurrence of any shocks; both sonic and Alf\'ven Mach numbers are less than unity \cite{Brown_psst}. As a result, the compression is quasi-static. Electron temperature, $T_e$ has been reported earlier in SSX Taylor states \cite{Brown_psst}. We have found that $T_e$ is around $\sim10~eV$ and is much less temporally dynamic than $T_i$.  

The time variation of the magnetic field from the long $\dot{B}$ probe array confirms the presence of a relaxed helical structure inside the SFC \cite{gray_prl}. From the magnetic fluctuation spectrum of the relaxed Taylor state in the compression volume, we found that the spectrum is dominated by energy at low frequencies and long wavelengths. The spectral index is much steeper for the relaxed object ($\alpha < -4$) than that observed in turbulence studies during relaxation ($\alpha = -2.47$) \cite{Schaffner2014}. Our primary objective in the present work is to study the additional heating of the fully-relaxed Taylor state due to compression.

We extract a time trace of the dominant axial wavenumber of the magnetic probe data using wavelet analysis. Because the structure is well described by the Taylor equilibrium, we can map the axial wavenumber directly to the length of the object. We found that the observed pitch of the Taylor state from axial $\dot{B}$ probe array is consistent with earlier more detailed measurements \cite{gray_prl} which leads us to believe that the profiles are also consistent.
As the Taylor state reaches the SFC and stagnates, its kinetic energy transforms primarily into an increase in its thermal energy. 

\begin{figure}
\includegraphics[width=0.49\textwidth]{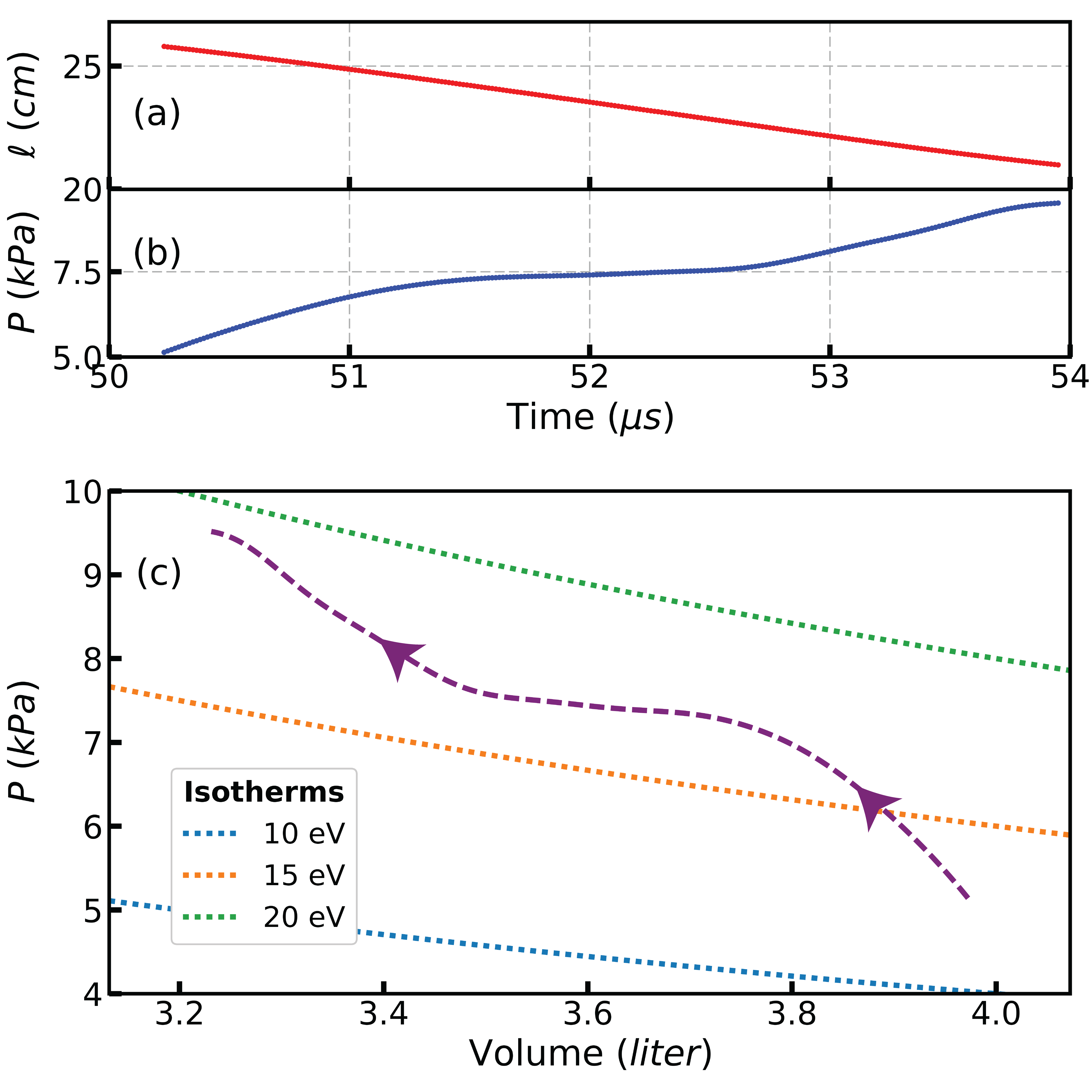}
\caption{\label{fig:PV_dia} a) A time trace (for the same shot as Fig.~\ref{fig:plasma_param}) of length of the Taylor state, b) an associated increase in plasma pressure and c) an excursion of the thermodynamic state of the object in a PV diagram for the compression process. Note that as the volume of the Taylor state decreases, the pressure shifts to higher isotherms.}
\end{figure}

Using time traces of Taylor state length and plasma pressure, we construct a PV diagram for the compression process as shown in Fig.~\ref{fig:PV_dia}.%
 The PV diagram clearly shows that as the Taylor state compresses, the plasma pressure increases, resulting in a shift to a higher isotherm in the PV diagram. In other words, compression and ion heating occur simultaneously and we call this a compressive heating event. In these experiments, we observed Taylor state length compression of up to $30\%$. 

After the identification of a compressive heating event, the equations of state (EOS) of the compressed plasma are computed to assess the thermodynamics of the magnetized plasma. Because SSX plasmas relax to an equilibrium described by MHD \cite{cothran_prl, gray_prl}, we might expect the thermodynamics to be described by the corresponding EOS. To check the validity of equations of state, hundreds of shots are recorded under a variety of gun parameters. In all these shots, compressive heating events are identified on a PV diagram. The compression events for most of the shots range from $2 - 3~ \mu s$ in duration. 

\begin{figure}
\includegraphics[width=0.52\textwidth]{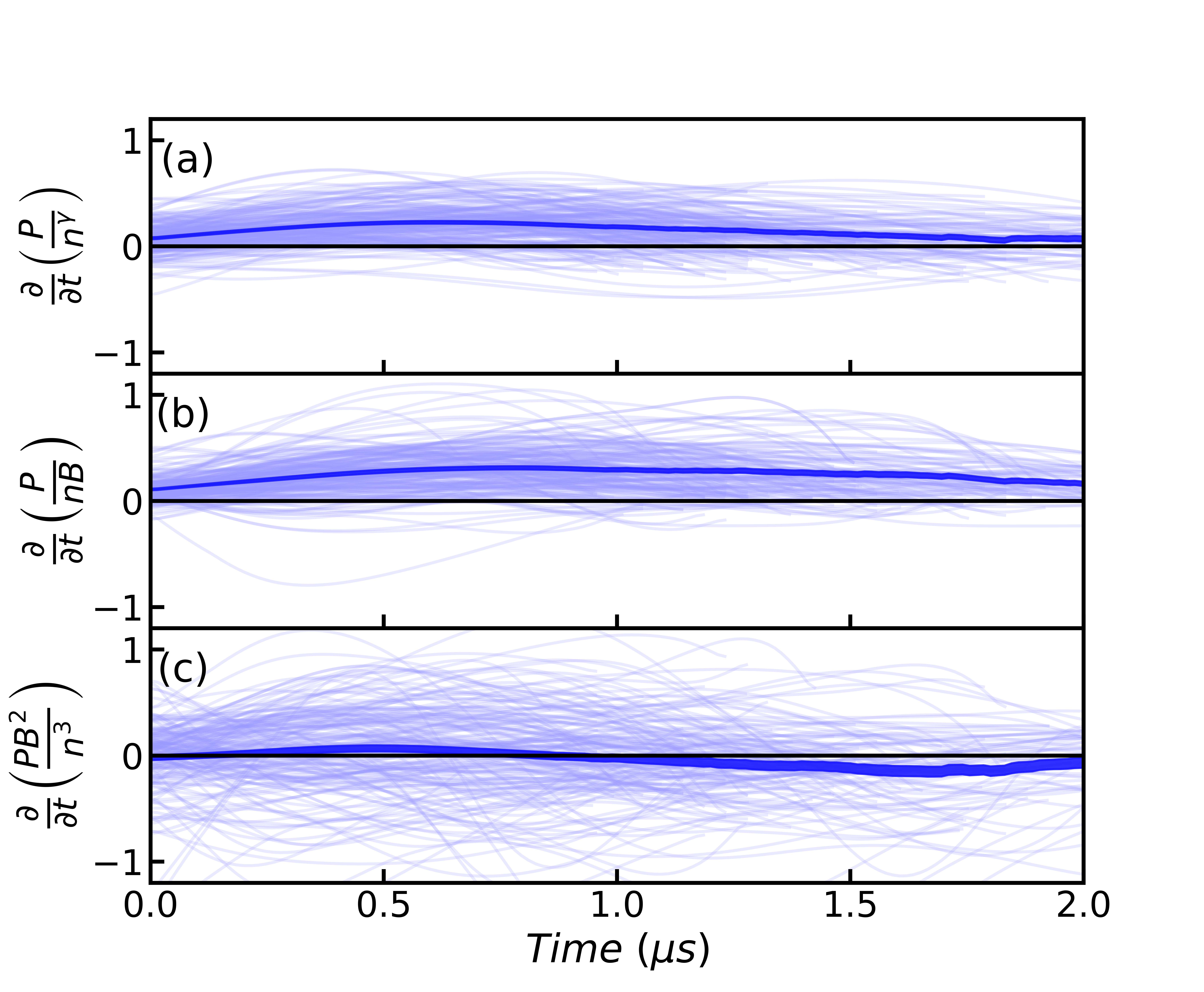}
\caption{\label{fig:eos} Statistical variation of three equations of state for 192 compression events (length contraction ranges from $10 - 30\%$): a) the magnetohydrodynamic equation of state for 3D compression ($\gamma = 5/3$), b) perpendicular, and c) parallel CGL equations of state as a function of time. The blue curve in each panel shows the standard error of the mean. Note that the MHD equation of state (a), and the perpendicular version of the CGL EOS (b) have nonzero time derivative. However, the parallel version of the CGL EOS (c) has nearly zero time derivative for most the compression time.}
\end{figure}

To determine the general behavior of the compressed states, we plot the EOS for all the events, as is shown in Fig.~\ref{fig:eos}.
Before taking the time derivative, quantities in each panel are normalized using their respective maxima such that the unit of each EOS in Fig.~\ref{fig:eos} is inverse of time. Since we measure ion temperature along a chord, our $T_{i}$ measurements are insensitive to the direction of magnetic field. We use total pressure to test the equations of state. In Fig.~\ref{fig:eos}, note that the mean time derivatives specified by Eq. \ref{mhd_eos} - \ref{<cgl_perp>}, are clearly nonzero, whereas the time derivative specified by Eq. \ref{<cgl_par>} remains nearly zero during compression. 

All the compression events used to construct Fig.~\ref{fig:eos} strictly satisfy the following criteria: ($i$) Length compression is greater than $10~\% $, ($ii$) compression event occurs for more than $1~\mu s$, and ($iii$) a transition from a lower to higher isotherm is demonstrated by the PV diagram, i.e., enough heating during compression. 

While strictly speaking, our measurements of temperature are insensitive to the direction of the magnetic field, it is striking that the parallel CGL EOS better characterizes our experiments. We hypothesize that as our spheromak expands immediately after formation and the structure unravels towards a Taylor state, the magnetic field magnitude drops by a factor of 10 causing a concomitant drop in $T_\bot$ due to conservation of magnetic moment. By the time the Taylor state enters the stagnation flux conserver, we suspect that most of the ion energy is in $T_\parallel$. The compression events in the stagnation flux conserver reduce the volume by only 10 - 30\%, which is not enough to lead to a considerable increase in $T_\bot$. The expansion of the solar wind may involve similar physics.

Impurities in the hydrogen plasma, through ionization and excitation states, may affect the $\gamma = 5/3$ assumption to MHD EOS. However, 
our past studies have shown that the impurity level in SSX is minimal 
\cite{Chaplin2009} and our plasma is optically thin so we do not expect the radiation to contribute to the EOS significantly. In any case, 
incorporating these effects is beyond the scope of this paper.

In summary, we present a PV analysis of the compression process of a magnetized, relaxed plasma leading to ion heating. The MHD EOS does not agree with experimental observations. Interestingly, the parallel component of the CGL EOS seems to model the average behavior of our plasma. 

In our future experiments, we wish to accelerate these Taylor states to higher velocities and then compress them to higher densities using pinch coils. In these experiments, we expect to observe higher compression. We are also preparing comparisons of our experiments with MHD simulations using the NIMROD code.

This work is supported by Department of Energy through ARPA-E ALPHA program and Office of Fusion Energy Sciences. We wish to acknowledge Simon Woodruff for general discussions, and Steve Palmer and Paul Jacobs for their technical support. 

\nocite{*}

\bibliography{Dr_Manjit_Kaur_revised}

\end{document}